\documentclass[11pt,fleqn]{article}
\usepackage{amssymb}

\newcommand{\half}{\mbox{$\frac{1}{2}$}}
\newcommand{\fslash}{\!\!\not\!}
\newcommand{\dslash}{\!\!\not\!\partial}
 
\begin{document}

\begin{table}
\begin{flushright}
IK--TUW 9810401
\end{flushright}
\end{table}

\title{Two-dimensional bosonization from \\
variable shifts in the path integral\thanks{Supported by Fonds zur
F\"{o}rderung der wissenschaftlichen Forschung, P11387--PHY}}

\author{Jan B. Thomassen\thanks{E-mail: {\tt
thomasse@kph.tuwien.ac.at}} \\
{\em Institut f\"{u}r Kernphysik, Technische Universit\"{a}t Wien} \\
{\em A--1040 Vienna, Austria}}

\maketitle

\begin{abstract}
A method to perform bosonization of a fermionic theory in $(1+1)$
dimensions in a path integral framework is developed. The method
relies exclusively on the path integral property of allowing variable
shifts, and does not depend on the explicit form of Greens
functions. Two examples, the Schwinger model and the massless Thirring
model, are worked out.

\vspace{\baselineskip}
\end{abstract}

\noindent
It is possible to argue that the problem of deriving the chiral
Lagrangian \cite{gasser} from QCD is a bosonization problem. This
motivates us to consider the simpler problem of bosonization in
$(1+1)$ dimensions. This is of course a well known problem with a long
history. A partial list of references is [2--14]. See
also the book \cite{abdalla}. Unfortunately, these approaches to
two-dimensional bosonization can either not handle more complicated
cases, like massive fermions, or they can only do so by comparing
explicit Greens functions in the fermionic and bosonic theories,
respectively. This includes the so-called ``path integral
approaches''. Thus they are not generalizable to four dimensions
where Greens functions can not in general be calculated. The need for
a path integral bosonization method which only utilizes manipulations
of the path integral itself is therefore apparent.

We find that a suitable starting point is the ``smooth bosonization
scheme'' of ref. \cite{damgaard}. We refer the reader to that paper
for details.

The generic problem which we consider is a massless fermion coupled to
an external vector field $V$. More complicated cases like a massive
fermion or non-Abelian bosonization are not studied in the present
paper, nor is a possible generalization to four dimensions.

The partition function is given by
\begin{eqnarray}
Z[V] & = & \int{\cal D}\psi{\cal D}\bar\psi
\exp i\int d^2x\bar\psi[i\dslash-\fslash V]\psi .
\end{eqnarray}
It is understood that the path integral is regularized in a vector
current conserving way. Following ref. \cite{damgaard}, we perform
first the arbitrary local chiral change of variables
\begin{eqnarray}
\psi(x) & = & e^{i\theta(x)\gamma_5}\chi(x), \hspace{2em}
\bar\psi(x) \; = \; \bar\chi(x)e^{i\theta(x)\gamma_5} .
\end{eqnarray}
The Lagrangian then becomes
\begin{eqnarray}
\label{rotated}
{\cal L} & = & \bar\chi[i\dslash-\dslash\theta\gamma_5
-\fslash V]\chi
+\frac{1}{2\pi}\partial_\mu\theta\partial^\mu\theta
+\frac{1}{\pi}\partial_\mu\theta\epsilon^{\mu\nu}V_\nu .
\end{eqnarray}
where the second and third terms is the contribution from the
Jacobian.

Next, we path integrate the ``chirally rotated'' partition function
over $\theta(x)$. Since we have only changed integration variables $Z$
doesn't depend on $\theta$. We can therefore perform this integration
without changing anything if we at the same time divide with the
(infinite) volume ${\cal N}=\int{\cal D}\theta$ of the $\theta$
function space.

In ref. \cite{damgaard} it is argued that we are dealing with a gauge
theory, invariant under the local transformations
\begin{eqnarray}
\nonumber
\chi(x) & \to & e^{i\alpha(x)\gamma_5}\chi(x), \hspace{2em}
\bar\chi(x) \; \to \; \bar\chi(x)e^{i\alpha(x)\gamma_5}, \\
\theta(x) & \to & \theta(x)-\alpha(x) .
\end{eqnarray}
However, we are not dealing with a gauge theory in the usual sense: It
is necessary to include the transformation of the integration measure
to get an invariant path integral. That is, the action itself is not
gauge invariant. For this reason gauge fixing is not necessary, and we
therefore leave the enlarged path integral intact (more on this
later). Thus, from this point on our procedure will differ from that
of ref. \cite{damgaard}.

We propose now to take advantage of having a bosonic path integration
variable at our disposal, in addition to the fermionic one. First, let
us note that we can immediately get back the original path integral by
applying the inverse rotation
\begin{eqnarray}
\chi & = & e^{-i\theta\gamma_5}\psi, \hspace{2em}
\bar\chi \; = \; \bar\psi e^{-i\theta\gamma_5}
\end{eqnarray}
to the fermion. But we can also ask if there is a {\em bosonic} change
of variables which decouples the fermion from $\theta$ and the
external field $V_\mu$:
\begin{eqnarray}
\theta & = & \theta(\phi,\bar\chi,\chi).
\end{eqnarray}
One simple case of such a bosonic variable change is the shift
\begin{eqnarray}
\label{shift}
\theta & \to & \theta+f(\bar\chi,\chi),
\end{eqnarray}
where $f$ is a function that only depends on the fermions. We shall
see that a function $f$ with the desired properties can be found.

Under the shift (\ref{shift}), the Lagrangian goes into
\begin{eqnarray}
\nonumber
{\cal L} \; \to \; {\cal L}'
  & = & \bar\chi[i\dslash-\dslash\theta\gamma_5-\dslash f\gamma_5
-\fslash V]\chi \\
  & & \mbox{} + \frac{1}{2\pi}\partial_\mu\theta\partial^\mu\theta
+\frac{1}{\pi}\partial_\mu\theta\partial^\mu f
+\frac{1}{2\pi}\partial_\mu f\partial^\mu f \\
\nonumber
  & & \mbox{} +\frac{1}{\pi}\partial_\mu\theta\epsilon^{\mu\nu}V_\nu
+\frac{1}{\pi}\partial_\mu f\epsilon^{\mu\nu}V_\nu.
\end{eqnarray}
The requirement for $\theta$ to decouple from the fermion is
\begin{eqnarray}
-\partial_\mu\theta(\bar\chi\gamma^\mu\gamma_5\chi)
+ \frac{1}{\pi}\partial_\mu\theta\partial^\mu f & = & 0,
\end{eqnarray}
whose solution is
\begin{eqnarray}
f & = &
\frac{\pi}{\square}\partial_\mu(\bar\chi\gamma^\mu\gamma_5\chi) 
  \; = \; \frac{\pi}{\square}\partial_\mu j^\mu_5.
\end{eqnarray}
With this value of $f$, the Lagrangian becomes
\begin{eqnarray}
{\cal L}' & = & \bar\chi i\dslash\chi
- \half\pi j^\mu_5\frac{\partial_\mu\partial_\nu}{\square}j^\nu_5
+\frac{1}{2\pi} \partial_\mu\theta\partial^\mu\theta
+ \frac{1}{\pi}\partial_\mu\theta\epsilon^{\mu\nu}V_\nu.
\end{eqnarray}
(Here we have used that $V_\mu$ can be chosen to be a curl,
$V_\mu=\epsilon_{\mu\nu}\partial^\nu\beta$, without loss of
generality.) As promised, the fermion is now completely decoupled and
$V_\mu$ couples only to $\theta$.

Note, incidentally, that what we have done is to perform a
diagonalization of the Lagrangian (\ref{rotated}). Since we still only
have normal fermionic and bosonic kinetic terms, this means that the
propagator matrix has only trivial zero modes. This demonstrates that
there is no need for gauge fixing.

In order to demonstrate that this procedure corresponds, when
applicable, to known results, we give two examples.

\bigskip
\noindent
{\bf i) The Schwinger model}

We set $V_\mu=eA_\mu$ and choose Lorentz gauge, $\partial_\mu A^\mu=0$
(this is trivially satisfied for $A_\mu$ a curl:
$A_\mu=\epsilon_{\mu\nu}\partial^\nu\beta$). In this case the integration over
$\theta$ should be explicitly carried out.
\begin{eqnarray}
\nonumber
Z & = & \int{\cal D}A{\cal D}\psi{\cal D}\bar\psi
\exp i\int d^2x\left(-\mbox{$\frac{1}{4}$}F_{\mu\nu}^2
+\bar\psi[i\dslash-e\fslash A]\psi\right) \\
  & = & \int{\cal D}A{\cal D}\theta
\exp i\int d^2x\left(-\mbox{$\frac{1}{4}$}F_{\mu\nu}^2
+\frac{1}{2\pi}\partial_\mu\theta\partial^\mu\theta
+\frac{e}{\pi}\partial_\mu\theta\epsilon^{\mu\nu}A_\nu\right) \\
\nonumber
  & = & \int{\cal D}A
\exp i\int d^2x\left(-\mbox{$\frac{1}{4}$}F_{\mu\nu}^2
+\frac{e^2}{2\pi}A^2\right)
\end{eqnarray}
This is the usual result \cite{roskies}, with a ``photon'' mass
$m=e/\sqrt{\pi}$.

\bigskip
\noindent
{\bf ii) The massless Thirring model}

We will in this case derive the current bosonization rule. We have
\begin{eqnarray}
\nonumber
Z[V] & = & \int{\cal D}\psi{\cal D}\bar\psi
\exp i\int d^2x\left(\bar\psi[i\dslash-\fslash V]\psi
-\half g j_\mu j^\mu\right) \\
\nonumber
  & = & \int{\cal D}B{\cal D}\psi{\cal D}\bar\psi
\exp i\int d^2x\left(\bar\psi[i\dslash-\fslash V-\fslash B]\psi
+\frac{1}{2g}B^2\right) \\
  & = & \int{\cal D}B{\cal D}\theta
\exp i\int d^2x\left(\frac{1}{2\pi}\partial_\mu\theta\partial^\mu\theta
-\frac{1}{\pi}(V_\mu+B_\mu)\epsilon^{\mu\nu}\partial_\nu\theta
+\frac{1}{2g}B^2\right) \\
\nonumber
  & = & \int{\cal D}\varphi
\exp i\int d^2x\left(\half\partial_\mu\varphi\partial^\mu\varphi
+\frac{\beta}{2\pi}V_\mu\epsilon^{\mu\nu}\partial_\nu\varphi\right),
\end{eqnarray}
where we have rescaled $\theta$,
\begin{eqnarray}
\varphi & = & \sqrt{\frac{1}{\pi}\left(1+\frac{g}{\pi}\right)}\;\theta,
\end{eqnarray}
and introduced the parameter $\beta$ by
\begin{eqnarray}
\frac{4\pi}{\beta^2} & = & 1+\frac{g}{\pi}.
\end{eqnarray}
The standard bosonization rule
\begin{eqnarray}
\bar\psi\gamma^\mu\psi
  & = & -\frac{\beta}{2\pi}\epsilon^{\mu\nu}\partial_\nu\varphi
\end{eqnarray}
can now be verified by reading off how $V_\mu$ couples to the fermion
$\psi$ and the boson $\varphi$, respectively.

\bigskip
Finally, we will offer some speculation on possible generalizations of
this scheme. It is clear that we should be able to treat more
complicated cases, such as the massive case or the non-Abelian case,
before we could claim having a genuine path integral bosonization
procedure. Let us note that we have not exploited all the degrees of
freedom of the fermion. We have for example, the scale degree of
freedom. Consider a massive fermion for definiteness. A finite local
scale transformation of the fermion will produce a Jacobian and also
couplings to the bosonic ``dilaton field''. Perhaps it is possible to
find shifts of this dilaton in combination with shifts of $\theta$
that will provide a decoupling of the fermion also in this massive
case. If so, integrating out the dilaton would produce a bosonic
theory depending only on $\theta$.

\noindent
\paragraph{Acknowledgments} I would like to thank M. Faber,
A.N. Ivanov and P.H. Damgaard for discussions and comments on the
manuscript. I also thank H.B. Nielsen for ``selling me some of his
ideas'' and thereby introducing me to the subject.

\end{document}